# Spatial multiresolution analysis approach to identify crash hotspots and estimate crash risk

Samer W. Katicha[1], John Khoury[2], and Gerardo Flintsch[1,3]

[1] *Center for Sustainable Transportation Infrastructure, Virginia Tech Transportation Institute*
[2] *Department of Civil Engineering, Lebanese American University*
[3] *The Charles E. Via Jr. Department of Civil and Environmental Engineering, Virginia Tech*

**Abstract**

In this paper, the authors evaluate the performance of a spatial multiresolution analysis (SMA) method that behaves like a variable bandwidth kernel density estimation (KDE) method, for hazardous road segments identification (HRSI) and crash risk (expected number of crashes) estimation. The proposed SMA, is similar to the KDE method with the additional benefit of allowing for the bandwidth to be different at different road segments depending on how homogenous the segments are. Furthermore, the optimal bandwidth at each road segment is determined solely based on the data by minimizing an unbiased estimate of the mean square error. The authors compare the SMA method with the state of the practice crash analysis method, the empirical Bayes (EB) method, in terms of their HRSI ability and their ability to predict future crashes. The results indicate that the SMA may outperform the EB method, at least with the crash data of the entire Virginia interstate network used in this paper. The SMA is implemented in an Excel spreadsheet that is freely available for download.

**Introduction**

Hazardous road segments identification (HRSI), or hotspot identification, is an important aspect of highway safety that is still an active research area (Jia et al. 2018; Yu et al. 2014; Elvik 2007 & 2008; Anderson 2009; Cheng and Simon 2005 & 2008; Huang et al. 2009; Montella 2010; Qu and Meng 2014; Park et al. 2014; Fawcett et al. 2018). The most common approaches for HRSI are based on a Bayesian approach (empirical Bayes or full Bayes) (Hauer 1997; Hauer et al. 2002; Elvik 1997, 2007 and 2008; Huang et al. 2009; Lord and Park 2008; Miaou and Song 2005) or a spatial analysis approach (Flahaut et al. 2003; Anderson 2009; Yu et al. 2014; Xie and Yan 2008; Loo et al. 2011). Recently, Yu et al. (2014) found that the kernel density estimation (KDE) spatial analysis method for HRSI performed well when compared to the empirical Bayes (EB) method although the EB method was in general better. One possible reason for the good performance of spatial models is that in many cases, the spatial model can account for a large portion (59 to 88%) of the heterogeneous crash variation (Barua et al. 2016; Aguero-Valverde and Jovanis 2008 and 2010; El-Basyouny and Sayed 2009). An additional benefit of spatial analysis methods is that they are well suited for large network analysis compared to other methods that generally do consider area-wide risk factors and have the advantage of only requiring the crash data (Yu et al. 2014).

Defining the road segments is an important aspect of HRSI (Yu et al. 2014; Chung et al. 2009; Thomas 1996). Researchers have used segmentations based on a constant segments length or segmentations based on homogenous segments with variable lengths. In both cases, the segmentation relies heavily on subjective judgement (Yu et al. 2014; Thomas 1996). With the road segments defined, the KDE approach assumes that crashes occurring on neighboring segments are correlated and the crash risk, defined as the expected number of crashes (sometimes called expected average crash frequency per year), varies smoothly along the segments. The random fluctuations of the observed crash counts can be "smoothed out" with the use of KDE with the amount of smoothing depending on the selected KDE bandwidth. The bigger the selected bandwidth, the more smoothing is performed. Therefore, selecting a bandwidth that is appropriate to smooth out most of the random fluctuations while at the same time preserving the variations that are due to the true crash risk is the most important aspect of KDE (Yu et al. 2014; Xie and Yan 2008; Anderson 2009; Flahaut et al. 2003). The optimal bandwidth would in principle minimize the integrated mean square error (Yu et al. 2014). It is not clear in the traffic safety literature how this is performed with most researchers selecting a bandwidth based on personal judgement (Anderson 2009; Xie and Yan 2008; Chung et al. 2009) or based on the number of observation (Flahaut et al. 2003; Yu et al. 2014). For example, Anderson (2009) selected a bandwidth of 200m, Chung et al. (2009) selected 32 m, and Xi and Yan (2008) compared results at six different bandwidth ranging from 20 m to 2000 m without specifying an optimal one. In practice, the optimal bandwidth depends on the smoothness of the function being estimated (in this case the function being estimated is the crash risk as a function of the spatial variable) and can actually be different at different road segments. The Spatial Multiresolution Analysis (SMA) method developed in this paper allows for the bandwidth to be different at every road segment with the optimal bandwidth determined by minimizing an unbiased estimate of the mean square error that can be calculated based only on the crash data.

**Important Aspects of Spatial Crash Analysis**

*The Relationship between Segment Length and Kernel Bandwidth*

The last paragraph of the introduction discussed road segmentation and KDE bandwidth selection. In spatial crash analysis, these two factors have mostly been considered independent of each other. However, as argued in this section, these two factors are closely related to the point of almost being interchangeable.

To illustrate how segmentation and bandwidth are related, a simple one dimensional example of a single road with the crash data initially obtained for segments having a length of 0.1 miles is considered. This could be performed by mapping global positioning system (GPS) coordinates of crashes with the GPS coordinates of the mile markers of the road (note that the choice of 0.1 miles length in this example is arbitrary and the same discussion applies to 0.01 miles, 1 mile, or any other chosen length). At this stage, the focus is not on how to determine the appropriate segment length or KDE bandwidth but on showing the close relation between a chosen segment length in segmentation and a chosen bandwidth in KDE. Therefore, comparison of these two data processing alternatives is performed as follows:

- Alternative 1 – Data aggregation: aggregate data into 0.3 miles segments.
- Alternative 2 – KDE processing: process the data with a KDE having a rectangular window of 0.3 miles (this is the bandwidth; also this is basically a moving average, the simplest form of KDE).

For Alternative 1, the crash counts of every 3 adjacent 0.1-miles sections (without overlapping) are added to obtain the required data aggregation. For Alternative 2, the crash counts of every 3 adjacent 0.1-miles sections (with overlapping) are added and divided by 3 to obtain the KDE result. It is easy to notice that Alternative 1, data aggregation, can be obtained from Alternative 2, KDE processing; simply sample the KDE results at 0.3 miles interval and multiply them by 3. This shows that segment length in road segmentation is essentially equivalent to the bandwidth of the moving average KDE. Of course in KDE analysis, kernels other than the rectangular window (e.g. Gaussian kernel) can be used. However it is well known that the choice of kernel does not significantly impact the results of the analysis (Silverman, 2018; O'Sullivan and Wong, 2007; Flahaut et al. 2003; Yu et al. 2014; Bil et al. 2013). KDE methods with different kernel shapes are basically similar to a weighted moving average. Therefore, the bandwidth of these kernel is also closely related to the segmentation length. With this equivalence between segment length and KDE bandwidth, a variable segment length data analysis is similar to a KDE analysis with a variable bandwidth. Although the simple example that was presented is one dimensional, the same argument is valid in a network space (see Xie et al. 2008 and Okabe et al. 2009) and a similar argument could be made in a two dimensional space. The SMA method proposed in this paper is implemented in a network space.

*The Need for a Variable Bandwidth*

To motivate the proposed SMA method Figure 1 shows the crash count recorded at 0.1 miles sections over a three year period on interstate 64 east in Virginia. This example highlights typical spatial features of crashes. Most of interstate 64 goes through rural Virginia. These areas have relatively low crash counts which results in long relatively homogeneous crash risk sections. Along interstate 64 are the urban areas of Richmond and Hampton Roads. These are characterized by high crash counts and high variation in crash risk so that homogeneous section would be much shorter. Finally, a 0.1 mile section in the Hampton Roads area has 82 recorded crash counts and stands out alone from all other sections. This example highlights the need for different section lengths and/or different bandwidth size when analyzing crash data.

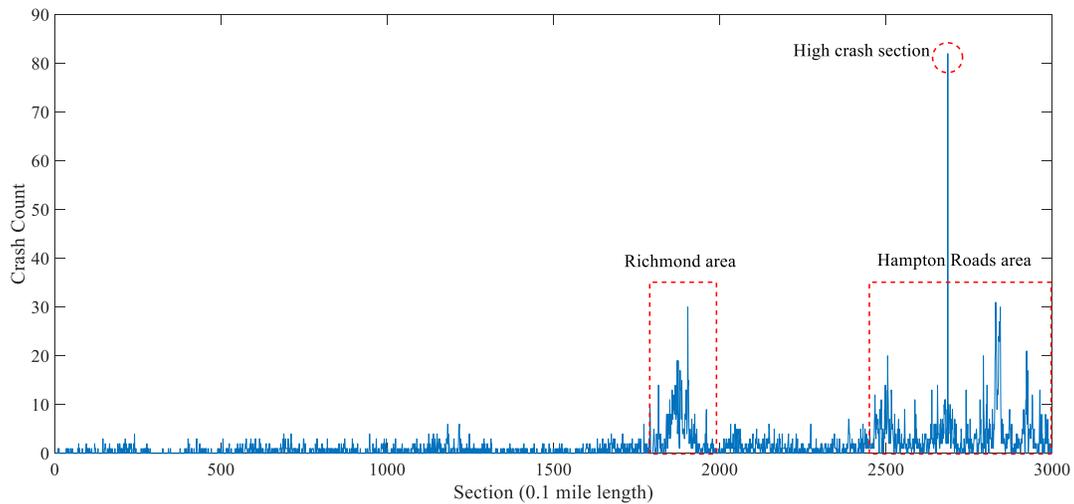

**Figure 1 – Crash count on interstate 64 east from 2014 to 2016**

Figure 2 shows a hypothetical crash risk along a road that has the same key features as those observed in Figure 1. A hypothetical model is used here because the performance of any analysis method can be easily evaluated. Crash count data were randomly generated and the crash risk was estimated from the crash counts using a KDE method with a Gaussian kernel and using the proposed SMA approach. For the KDE analysis, the optimal bandwidth was obtained by minimizing the mean square error with the true crash risk. For the SMA approach, the results were obtained without using the true crash risk and are based solely on the generated crash counts. Figure 3 shows the estimated crash risk from each analysis method. The SMA estimate is able to preserve the key features of the true crash risk. The section with the very high crash risk is accurately estimated and a smooth estimate is obtained at the locations where the crash risk is relatively homogeneous. The KDE estimate does not preserve the features of the true crash risk. The section with the very high crash risk is estimated as having a much lower crash risk (too much smoothing). The locations where the crash risk is relatively homogeneous are estimated as being relatively highly variable (not enough smoothing). This shows that a single bandwidth will not in general be adequate to analyze crash data and a variable bandwidth spatial analysis method is needed.

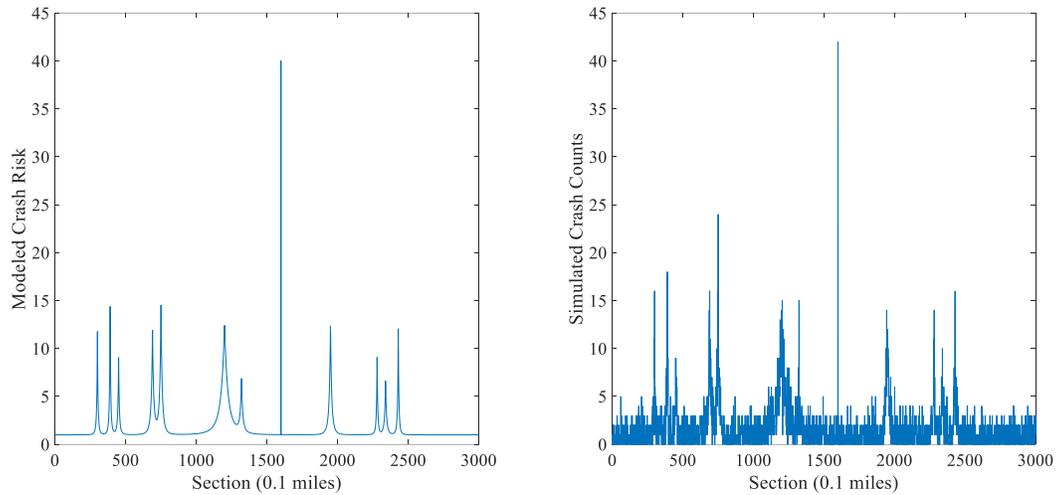

**Figure 2 – Crash risk model (left) and randomly generated crash counts (right)**

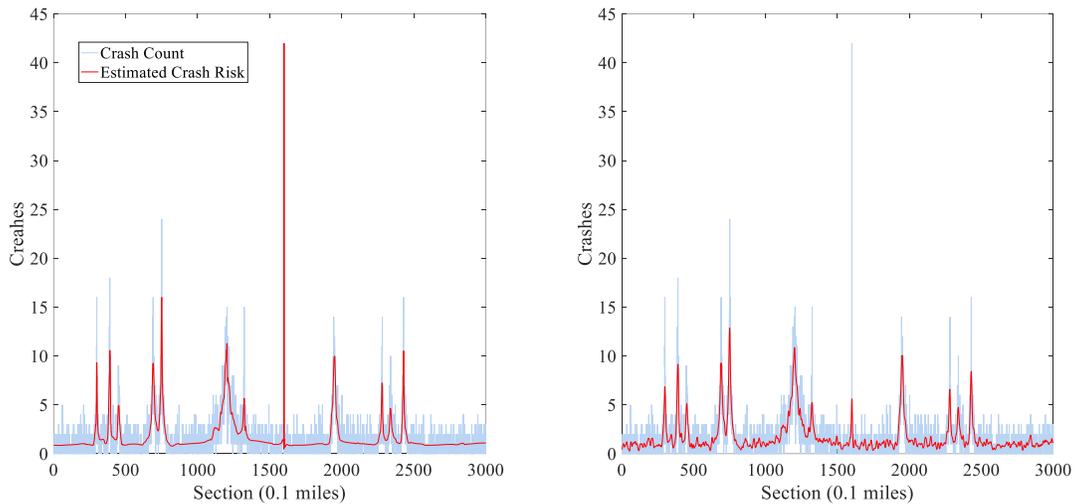

**Figure 3 – Estimated crash risk with the SMA (Left) and the KDE (Right) methods**

## Methodology

This section gives an overview of the EB method, presents the proposed SMA method and then presents the statistical tests used to evaluate and compare the two methods.

### *Empirical Bayes (EB) Method*

Bayesian methods, whether empirical or full Bayes, are the most widely used methods for crash data analysis (Hauer et al. 2002, Mannering et al. 2014, Elvik 2008b, Persaud et al. 2010). The EB method is based on the Safety Performance Function (SPF). According to the Highway Safety Manual (AASHTO 2010, page G-13) an SPF is "… an equation used to estimate or predict the expected average crash frequency per year at a location as a function of traffic volume and in some cases roadway or intersection characteristics (e.g., number of lanes, traffic control, or type of median)." The development of an SPF is generally performed using negative binomial regression. This assumes that the crash risk (which in our terminology is the same as "the expected average crash frequency per year"), $\lambda$, for sections having similar characteristics is related to those characteristics, as follows:

$$\lambda = \exp\left(\sum_{j=1}^{k} \beta_j X_j\right) \times \varepsilon = \mu \times \varepsilon \qquad (1)$$

$$\mu = \exp\left(\sum_{j=1}^{k} \beta_j X_j\right) \qquad (2)$$

Where,

$k$ = number of characteristics considered in the model (including intercept)

$X_j$ = j$^{th}$ characteristic

$\beta_j$ = regression coefficient for the j$^{th}$ characteristic

$\varepsilon$ = error terms assumed to be gamma distributed with mean 1

Though some section characteristics are included in the regression model, the error term accounts for the fact that the crash risk is also related to other factors not included in the regression that are considered to be adequately represented by a gamma distribution. Therefore, sections having the same characteristics considered in the regression model will have different crash risk. Crash counts are assumed to be generated from a Poisson process, as follows:

$$Y \sim P(Y = Z | \lambda) = \frac{\lambda^Z \exp(\lambda)}{Z!} \qquad (3)$$

Where,

$Y$ = crash count

$P(Y = Z | \lambda)$ = the probability of $Y$ taking the value $Z$ given the crash risk is $\lambda$

Negative binomial regression uses the crash counts $Y$ to estimate $\mu$ and the variance of $\varepsilon$ expressed in terms of the overdispersion. The overdispersion is related to the variance of the gamma distribution and the variance of the model error, as follows:

$$\sigma^2_{Gamma} = \phi \mu^2 \qquad (4)$$

$$\sigma^2_{Model\ Error} = \mu + \phi \mu^2 \qquad (5)$$

The estimated regression parameters, $\beta_j$, allow us to obtain an estimate $\hat{\mu}$ for $\mu$, while $\sigma^2$ can be estimated from the residuals making it possible to estimate the overdispersion $\phi$. The crash risk is estimate using the EB approach as a weighted average of regression model and the crash counts as follows:

$$\hat{\lambda} = \left(1 - \frac{\hat{\phi}\hat{\mu}}{1 + \hat{\phi}\hat{\mu}}\right)\hat{\mu} + \frac{\hat{\phi}\hat{\mu}}{1 + \hat{\phi}\hat{\mu}} Y \qquad (6)$$

The hat on the parameters is used to stress that the values used are estimated from the data.

*Spatial Multiresolution Analysis (SMA) Method*

The SMA method is similar to the (network) KDE method with the additional feature of allowing different bandwidths at different road sections. Furthermore, the SMA method incorporates an approach to determine the optimal bandwidth at each road section based on minimization of an unbiased estimate of the mean square error between the estimated crash risk and the true unknown crash risk. A self-contained Matlab implementation of the SMA method is given in the Appendix.

*SMA Method*

The SMA is based on a wavelet analysis method that uses the Haar wavelet. However, it is presented here as an extension of the moving average KDE (or as an extension of road segmentation) without the need to explain wavelet analysis or the discrete wavelet transform. For more details on the theoretical aspects of the SMA approach, the reader is referred to Katicha et al. (2018) and the references therein.

Suppose the crash data is recorded for 0.1-mile sections and consider two adjacent sections with crash counts $y_i$ and $y_{i+1}$. The two 0.1-mile sections can be combined into a 0.2-miles section having a crash count $s_i = y_i + y_{i+1}$. Similarly a two points moving average will give $s_i/2 = (y_i + y_{i+1})/2$. The remaining pairwise adjacent sections can also be combined to give $s_{i+2} = y_{i+2} + y_{i+3}$, $s_{i+4} = y_{i+4} + y_{i+5}$, ... (note that in the moving average case odd increments are also calculated $s_{i+1}/2 = (y_{i+1} + y_{i+2})/2$, $s_{i+3}/2 = (y_{i+3} + y_{i+4})/2$, ...). Combining the 0.1 mile sections into 0.2 mile sections results in half the amount of data. Furthermore, the information about the 0.1 mile sections is lost (the data for 0.1 mile sections cannot be obtained from the 0.2 mile sections). To preserve the information, the crash count difference between two adjacent 0.1 mile sections is calculated as follows: $d_i = y_i - y_{i+1}$, $d_{i+2} = y_{i+2} - y_{i+3}$, $d_{i+4} = y_{i+4} - y_{i+5}$, ... The counts at each 0.1 mile section can be recalculated from the sums and differences as follows: $y_i = (s_i + d_i)/2$ and $y_{i+1} = (s_i - d_i)/2$. Note that if $d_i$ is set to zero, then instead of obtaining the original crash counts, the (moving) average crash count $(s_i + 0)/2 = s_i/2$ is obtained, showing that combining two 0.1-mile sections into one 0.2-mile section is equivalent to setting $d_i = 0$. If the crash risk, $\lambda_i$ and $\lambda_{i+1}$ at adjacent sections are similar, then it would be beneficial to consider the two sections as homogenous ($\lambda_i \approx \lambda_{i+1}$) and combine them to reduce the variability of the crash counts. If on the other hand $\lambda_i$ and $\lambda_{i+1}$ are very different, then it would be better not to combine the two sections. The true risk, $\lambda_i$ are not known however, small $d_i$ values (in absolute value) suggest that the sections have similar risks and therefore should be combined by setting $d_i = 0$ while large $d_i$ values suggest that the sections have different risks and should not be combined. A threshold $t$ is used to separate what is small from what is large (the next section explains how an optimal threshold $t$ can be determined). Because some $d_i$ values (the ones less than or equal to $t$ in absolute value) are set to zero, while others are not, the thresholding procedure will result in two different section lengths. Locations where $d_i$ is set to zero, the section length effectively becomes 0.2-mile whereas locations where $d_i$ is not set to zero, the section length remains 0.1-mile (the original length at which the data was acquired). The act of thresholding the $d_i$ values has effectively resulted in sections of different length which is equivalent to having performed averaging with two different window sizes (0.1 and 0.2 miles).

The $s_i$ values obtained by combining two 0.1-mile sections represent 0.2-mile sections. These 0.2-mile sections can again be analyzed following the same approach used to analyze the 0.1-mile sections. This is done by combining the data into 0.4-mile sections and calculating $2s_i = s_i + s_{i+2}$ and $ds_i = s_i - s_{i+2}$. A new threshold $t$ is used for $ds_i$ so that locations where $ds_i$ is set to zero will result in section lengths of 0.4 mile. This process can be repeated by doubling the section length at each iteration until the largest section length, that is less than or equal to the total length of the road, is reached.

Table 1 shows the process of obtaining the sums shown in the second column and the differences shown in the third column for a 100-miles road starting from 0.1-mile sections (This decomposition process is basically the discrete Haar wavelet transform). Only, the differences at every aggregation length (resolution) and the last sum at the largest aggregation length, $9s_i$, are needed to be able to recover the original data (the 0.1-mile section length data). The sums at every aggregation length can be obtained by combining the last sum with all the differences obtained at the aggregation level of interest and higher. For example, $8s_i$ can be obtained by combining $9s_i$ and $d8s_i$, and $6s_i$ can be obtained by combining $9s_i$, $d8s_i$, $d7s_i$ and $d6s_i$. Combining $9s_i$ with all the differences gives the original 0.1-mile data $y_i$. The fourth column in Table 1 shows thresholded differences with the appropriate threshold at each aggregation length. Using $9s_i$ and the thresholded differences (instead of the differences) at all aggregation lengths an estimate of the crash risk, denoted by $\hat{\lambda}_i$, is obtained at each 0.1-mile section. The resulting estimate would have different averaging lengths because thresholded differences are used. For example, at a specific location, all differences from $d_i$ up to and including $d2s_i$ could be set to zero, while at another location, all differences from $d_i$ up to and including $d6s_i$ could be set to zero. At the former location crash counts are essentially averaged over a 0.8-mile section, while at the latter location, crash counts are averaged over a 12.8-miles section. This shows that the approach results in different averaging lengths at different locations.

**Table 1 – Decomposition process of a 100-miles roadway section**

| Section Length (miles) | Sum | Difference | Thresholded Difference | Estimated Risk |
|---|---|---|---|---|
| 0.1 | $y_i$ | $d_i$ | $td_i$ | $\hat{\lambda}_i$ |
| 0.2 | $s_i$ | $ds_i$ | $tds_i$ | $ts_i$ |
| 0.4 | $2s_i$ | $d2s_i$ | $td2s_i$ | $t2s_i$ |
| 0.8 | $3s_i$ | $d3s_i$ | $td3s_i$ | $t3sr_i$ |
| 1.6 | $4s_i$ | $d4s_i$ | $td4s_i$ | $t4s_i$ |
| 3.2 | $5s_i$ | $d5s_i$ | $td5s_i$ | $t5s_i$ |
| 6.4 | $6s_i$ | $d6s_i$ | $td6s_i$ | $t6s_i$ |
| 12.8 | $7s_i$ | $d7s_i$ | $td7s_i$ | $t7s_i$ |
| 25.6 | $8s_i$ | $d8s_i$ | $td8s_i$ | $t8s_i$ |
| 51.2 | $9s_i$ | | | |

*Determining the Optimal Threshold at Each Averaging Length*

Using an optimal threshold at each averaging length is very important for the SMA approach to produce good results. Assume the true crash risk at each location is known. Using that number, the true differences at each averaging length can be easily calculated. Denote these differences at the smallest scale by $d\lambda_i$. If the $d\lambda_i$ were kwon then they can be used to determine the optimal threshold that minimizes the mean square error between $td_i$ and $d\lambda_i$. Because $d\lambda_i$ are actually not known, the true mean square error cannot be calculated. However, for Poisson data, an unbiased estimate of the mean square error, called PURE (**P**oisson's **U**nbiased **R**isk **E**stimate), can be calculated solely based on the data (i.e. $d_i$ and $td_i$). For the calculated differences at any aggregation length, PURE can be calculated as follows:

$$\text{PURE}(\text{th}) = \|\theta(\mathbf{d},\mathbf{s},\text{th})\|_2^2 + \|\mathbf{s}\|_1 + 2\mathbf{d}^\text{T}\theta(\mathbf{d},\mathbf{s},\text{th}) - (\mathbf{d}+\mathbf{s})^\text{T}\theta(\mathbf{d}-\mathbf{1},\mathbf{s}-\mathbf{1},\text{th}) \\ + (\mathbf{s}-\mathbf{d})^\text{T}\theta(\mathbf{d}+\mathbf{1},\mathbf{s}-\mathbf{1},\text{th}) \tag{7}$$

Where **d** and **s** are the vectors of differences and sums and th is the threshold. $\theta$ is a function of **d** and **s** and the threshold th related to the thresholding function as follows: $\theta(\mathbf{d}, \mathbf{s}, \text{th}) = \varphi(\mathbf{d}, \mathbf{s}, \text{th}) - \mathbf{d}$, where $\varphi(\mathbf{d}, \mathbf{s}, \text{th})$ is the thresholding function used. To determine the optimal threshold, PURE is calculated for a set of thresholds th and the one that gives the lowest PURE is select. Refer to Katicha and Flintsch (2018) for an illustration of how good PURE is as an unbiased estimate of the mean square error. Luisier et al. (2010) and Hirakawa et al. (2012) present a proof that PURE is an unbiased estimate of the mean square error.

*Details Regarding the Thresholding Function*

In the SMA approach differences less than or equal to the threshold (in absolute value) are set to zero while the larger ones are not. An obvious choice for differences larger than the threshold is to keep them unmodified. This approach is referred to as hard thresholding (there is a hard cutoff), which is too abrupt and can lead to numerical instabilities in the calculation of PURE. An alternative to hard thresholding is to reduce the differences that are larger (in absolute value) than the threshold by a value equal to the threshold. This approach is known as soft thresholding and is more stable for the calculation of PURE. However, soft thresholding, reduces all differences by a value equal to the threshold, which in general, is not desirable for very large differences. Even with that drawback, soft thresholding is generally better than hard thresholding. It is possible to use another thresholding function that is continuous like the soft thresholding function but also converges to the hard thresholding function for high values of the differences. This function will result in numerically stable PURE like soft thresholding without the drawback of reducing very large differences. Figure 4 shows the hard thresholding function, the soft thresholding function, and a continuous thresholding function that converges to the hard thresholding function for large differences. In the figure, a threshold of 5 is used for illustration. In the implementation, a thresholding function similar to the one shown in Figure 4c is used.

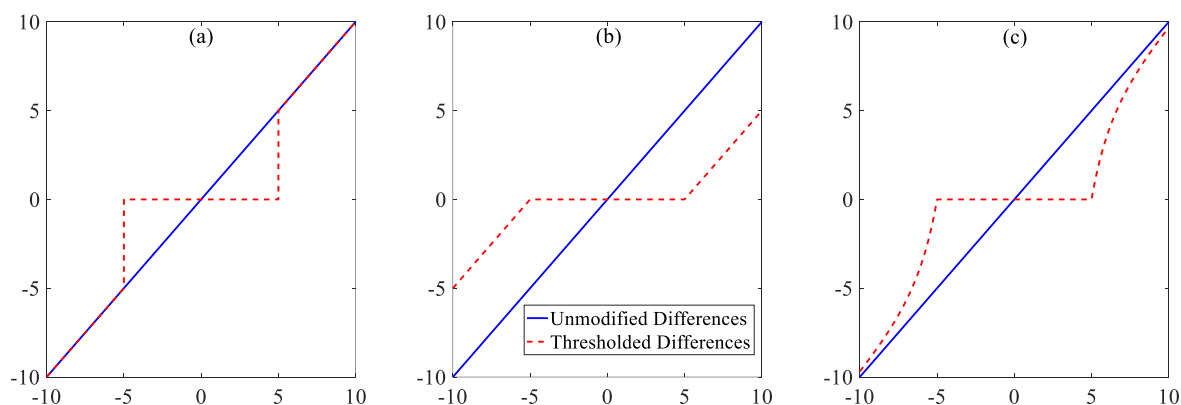

**Figure 4 - Example of thresholding functions: (a) hard thresholding, (b) soft thresholding, and (c) continuous thresholding converging to hard thresholding for large differences**

*Quantitative Evaluation of Analysis Methods*

Four quantitative evaluation tests are used to compare the SMA method with the EB method, and the Count method (i.e. just using the crash counts as estimate of the crash risk). Three tests, which were used by Yu et al. (2014), are used for HRSI evaluation. The fourth test is the mean square prediction error (MSPE) and is used to estimate the accuracy of crash prediction (for all crashes). The four tests are performed on 0.1 mile road sections (the section length at which the data was obtained) and are presented below.

*Segment Consistency Test*

In the segment consistency test (SCT), the hazardous sections are identified by each method in a given time period (usually a year). The average crash counts on these identified sections is calculated for the following time period to give the SCT for each method. The method with the higher SCT is the one that is considered better for HRSI evaluation. SCT can be calculated for different thresholds $\alpha$ used to identify hazardous road sections. For example an $\alpha = 0.05$ means that the top 5% of road sections are considered hazardous sections.

*Method Consistency Test*

In the method consistency test (MCT), the hazardous sections are identified by each method for two consecutive time periods. The MCT is the percentage of road sections that are identified in both time periods for a specific threshold $\alpha$.

*False Positive Test*

As described by Yu et al. (2014), the false positive rate is the proportion of safe road sections that are identified as hazardous sections. Yu et al. (2014) used the false identification test which also measures the false negative rate, where the false negative rate is defined as the proportion of hazardous road sections that are identified as safe sections. It can be shown that the method with the lowest false negative rate will also have the lowest false positive rate and lowest false identification. Therefore, only the false positive rate was calculated. Because the true hazardous road segments are not known a priori, Yu et al. (2014), used two versions of false identification test that they denoted by FIT-I and FIT-II. For the FIT-I approach they estimated a negative binomial crash predictive model using 10 years of data that they used as the reference to establish the true hazardous sections. This probably results in a bias in favor of the EB method for the following two reasons:

1. The EB method also uses a negative binomial model crash predictive and therefore should be expected to be more compatible with the results of the reference crash predictive model.
2. The reference model uses crash data from 2001 to 2010. Two models for the EB method evaluation were developed with data from 2005 to 2007 and data from 2008 to 2010. These two time periods are within the time period used for the reference model and therefore could further bias the results in favor of the EB method.

For this reason the data used as a reference for the false positive test was from a different time period than that used to evaluate the EB and SMA methods. Furthermore, three types of references are established and used to evaluate the methods. These are:

1. No modeling: in this case, the crash counts are used as the reference. This is similar to the method of FIT-II used by Yu et al. (2014).
2. EB model: this is the approach used by Yu et al. (2014) with FIT-I however, unlike in Yu et al. (2014) the dataset used to obtain the model does not overlap with the dataset used to evaluate the analysis methods.
3. SMA model: in this case, the results of the SMA on the independent dataset are used as the reference.

*Mean Square Prediction Error Test (MSPE)*

In the mean square prediction error (MSPE) test, the crash risk is identified by each method at every 0.1-mile section in a given time period (usually a year). The difference between the estimated crash risk and the crash count at every section in the following time period is used to calculate the MSPE error as follows:

$$MSPE = \frac{1}{N} \sum_{i=1}^{N} \left( Y_i^{j+1} - M_i^j \right)^2 \qquad (8)$$

Where,

$Y_i^{j+1}$ = crash count at section $i$ in time period $j+1$

$M_i^j$ = model estimate at section $i$ in time period $j$. The model can be the crash counts, the EB model, or the SMA model

$N$ = total number of observations

**Collected Crash Data**

The 2014, 2015 and 2016 crash data for the entire State of Virginia interstate network (2,236 directional miles) are used in the analysis. In each year, about 65% of the sections did not have a recorded crash (i.e. crash count of 0). Figure 5 shows a map of Virginia with the interstate network and reported crashes color coded for low crash count (green), medium crash count (yellow) and high crash count (red). Low crash counts are generally observed in rural areas, medium crash counts are generally observed in urban areas, and high crash counts are observed in major population areas (Northern Virginia, Richmond, and Tidewater area). Figure 6 shows the observed crash counts in 2014 for all interstate roads (both directions). For the EB model, the following variables were included: traffic, ramp, number of lanes, urban/rural, grade, horizontal curvature, vertical curvature, speed limit, and road identification.

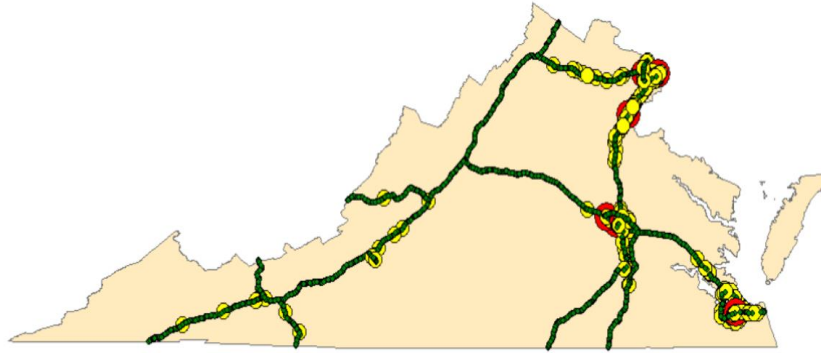

**Figure 5 Map of Virginia showing the interstate network and recorded crashes in 2014.**

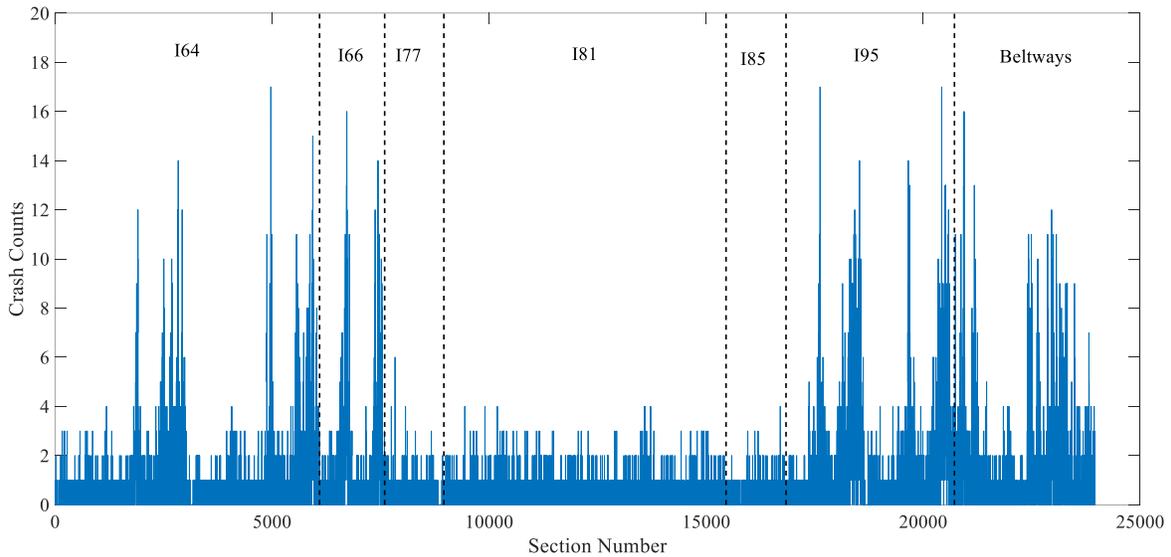

**Figure 6 - Observed crash counts in 2014 on every 0.1-mile section of the Virginia interstate roads**

## Results and Discussion

### *SPF Approach*

Table 2 show the fitted model parameters. Figure 7 shows the estimated crash risk (expected crashes) obtained with the EB approach along with the crash counts. It can be seen that the estimated crash risk follow the overall trends of the crash counts but are less variable.

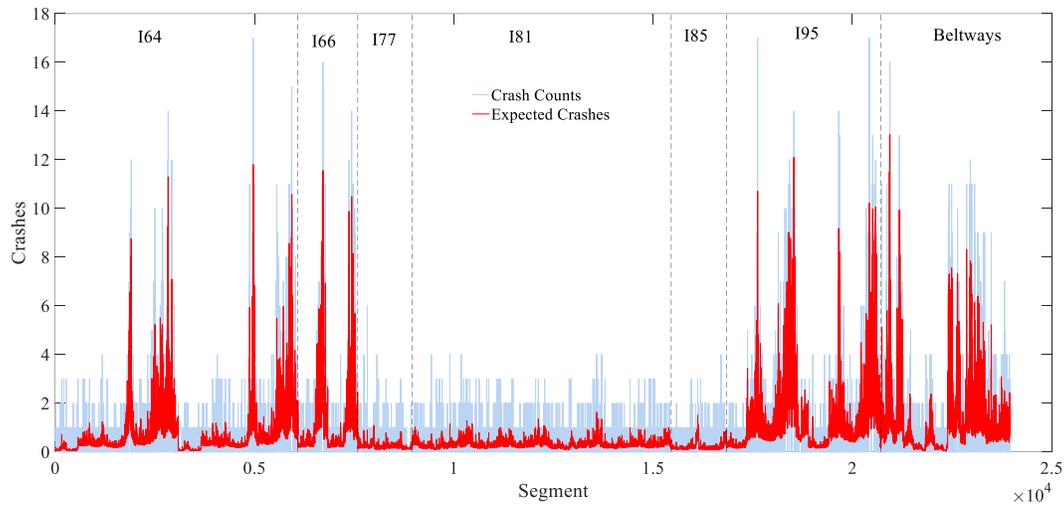

**Figure 7 - 2014 Crash counts and estimated crash risk using the EB method**

**Table 2 – Results of regression model**

| Parameters | E | Std. | $P>\chi^2$ |
|---|---|---|---|
| Overdispersion | 0.6605 | 0.0031 | |
| Constant | - | - | - |
| ln(AADT) | 1.0840 | 0.02166 | <0.0001 |
| Ramp | 0.0137 | 0.00077 | <0.0001 |
| Number of Lanes | 0.1792 | 0.01006 | <0.0001 |
| Urban | 0.3504 | 0.02226 | <0.0001 |
| Grade | -0.0160 | 0.00576 | 0.0054 |
| Horizontal Curvature | 0.1570 | 0.01650 | <0.00001 |
| Vertical Curvature | -0.0002 | 0.00004 | <0.00001 |
| Speed Limit | 0.0023 | 0.00068 | 0.0007 |
| Route ID | - | - | - |
| Log likelihood | -22,877 | | |
| AIC | 45,802 | | |
| BIC | 45,996 | | |

*Spatial Multiresolution Analysis (SMA) Approach*

Figure 8 shows the estimated crash risk (expected crashes) along with the crash counts. Compared to the results shown in Figure 7, the results shown in Figure 8 are much smoother. The variable bandwidth feature of the SMA that results in different averaging at different locations is also evident. Interstate 77, 81, and 85 are mostly in rural areas which show very little spatial variation and the results show that a relatively large averaging length is used to estimate the crash risk. On the other hand, Interstate 64, 66, and 95 have much more spatial variation especially around the large urban areas and the results show that a much smaller averaging length is used to preserve the spatial variation. Figure 9 shows the bandwidth at each 0.1 mile road segment. Because the SMA is similar to a variable bandwidth KDE with a rectangular window, the bandwidth length could also be interpreted at aggregation length. The figure shows that most small bandwidths occur at urban areas and most large bandwidths occur at rural areas. Figure 10 shows the proportion segments in each bandwidth size. The majority of road segments (~70%) had a bandwidth size between 1.6 miles and 12.8 miles and very few of the road segments (less than 0.5%) had a bandwidth size of 0.1 mile, which is the resolution at which the data was collected.

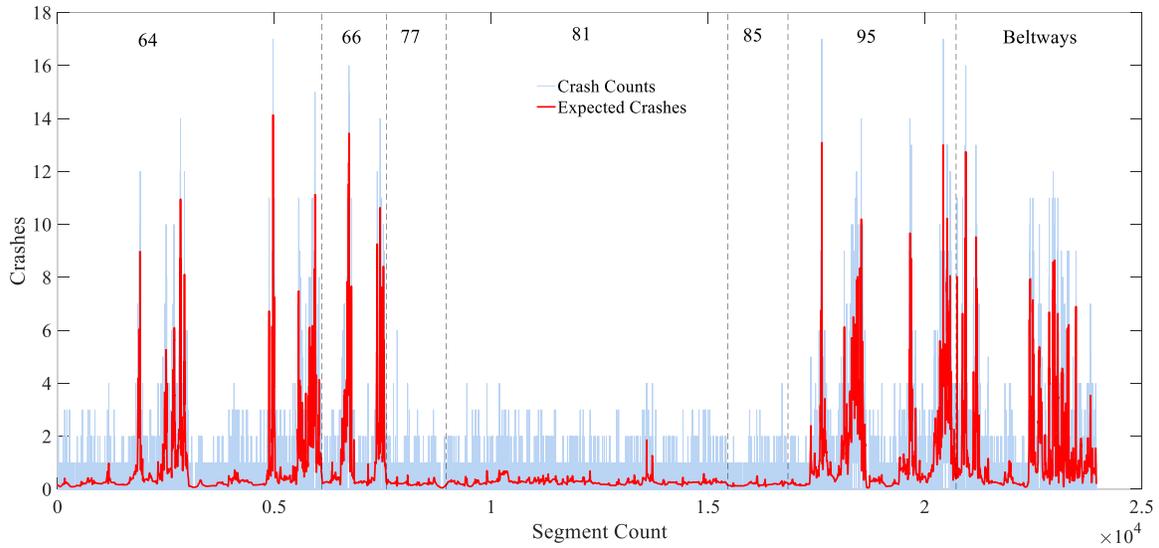

**Figure 8 – 2014 Crash counts and estimated crash risk using the SMA**

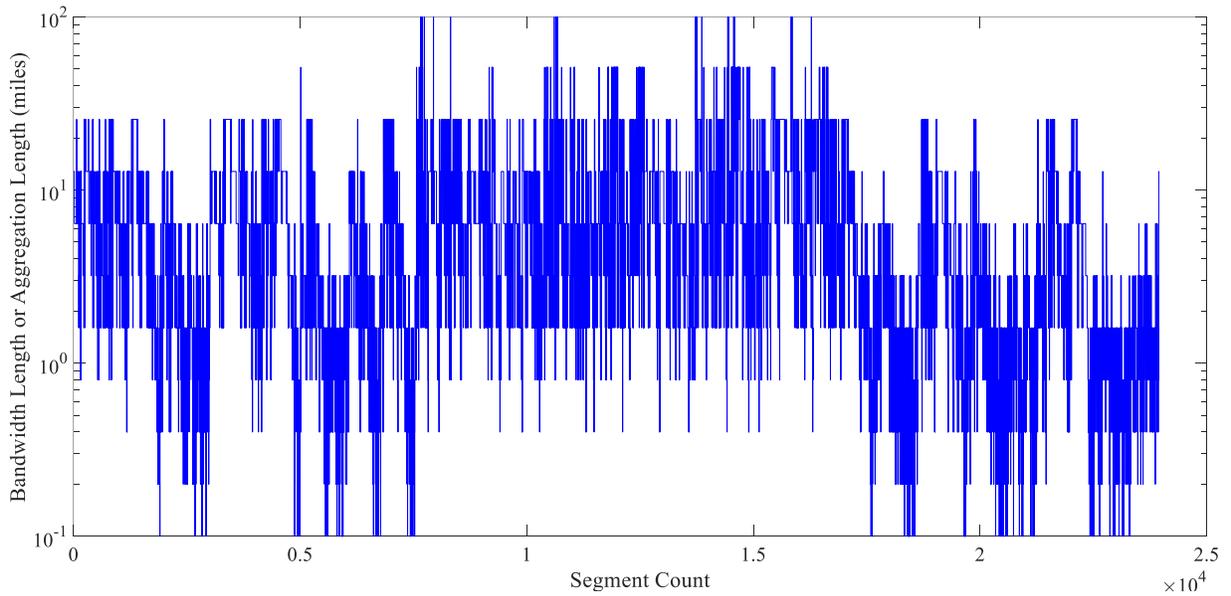

**Figure 9 – Bandwidth of the SMA at each road segment**

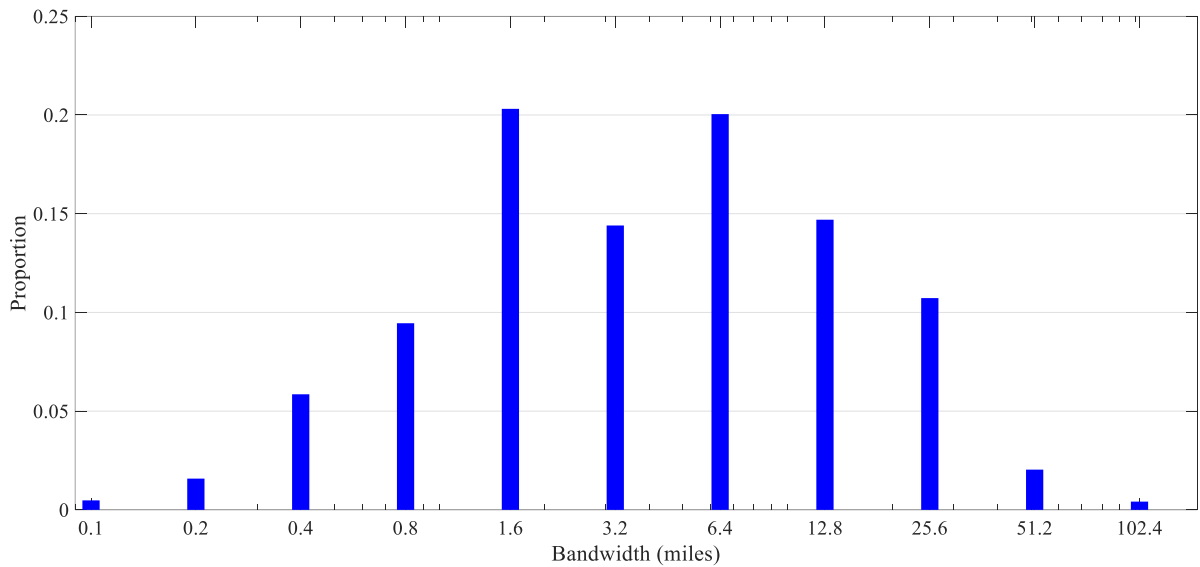

**Figure 10 – Proportion of segments averaged at a specific bandwidth size**

*Discussion*

Table 3, Table 4, Table 5 and show the results of SCT, MCT, and FP and MSPE, respectively. For SCT and MCT, higher values indicate better performance while for FP and MSE, lower values indicate better performance. In all evaluated cases, the SMA had the best performance. Except for a couple of FP cases with the crash count data with α set to 1%, the EB method performance was second. In general, the count approach performed by far the worse except for the FP with α set to 1%. The results of SCT are in general agreement with the results obtained by Yu et al. (2014) who found that the KDE approach generally performed the best (the SMA which is similar to a variable bandwidth KDE). The results of MCT generally contradict the results obtained by Yu et al. (2014) who found the spatial analysis methods performing the worse. However, the authors argued that segmentations could have the effect of reducing the performance of the spatial analysis methods compared to the conventional methods. In SMA, the starting segments for all methods was set to 0.1-mile and therefore the effect of segmentation is the same on all evaluated method. The FP results should in general be comparable to the false identification results of Yu et al. (2014). Yu et al. (2014) found that the EB method performed the best followed by the KDE method. In this paper, the SMA method performed best followed by the EB method. The SMA is an improvement over the KDE in terms of allowing a variable bandwidth and using PURE to optimize the bandwidth at each location. These features are improvement on the KDE approach and could have contributed to the SMA performing better than the EB. Furthermore, the analysis of the an entire road network, the interstate network, is well suited for spatial analysis methods as suggested by Yu et al. (2014). Finally, the EB results are specific to the SPF used. Additional significant variables could be included in the SPF which could potentially make the EB results closer or even better than those obtained with the SMA. However, obtaining data for additional variables on the entire interstate network could be difficult and time consuming especially when the benefits of obtaining the data are not really known beforehand.

**Table 3 Segment Consistency**

| Method | Performance | | | |
|---|---|---|---|---|
| | Top 1% | Top 2.5% | Top 5% | Top 10% |
| Count | 6.3000 | 5.0868 | 4.1352 | 3.0759 |
| EB | 6.3583 | 5.1803 | 4.2354 | 3.2706 |
| SMA | **6.5292** | **5.3823** | **4.4533** | **3.3925** |

**Table 4 Method Consistency**

| Method | Performance | | | |
|---|---|---|---|---|
| | Top 1% | Top 2.5% | Top 5% | Top 10% |
| Count | 34.17% | 42.74% | 47.60% | 51.77% |
| EB | 38.75% | 48.08% | 56.51% | 65.55% |
| SMA | **44.58%** | **55.43%** | **65.69%** | **75.37%** |

Because of the strong evidence pointing to the presence of spatial correlation in crash data, researchers have more recently started to use Bayesian models that incorporate spatial as well as temporal effects in HRSI (e.g. Cheng et al. 2017, Huang et al. 2016, Dong et al. 2016). In general, these models have performed better than Bayesian models that do not incorporate spatial and/or temporal effects. One limitation of these Bayesian spatiotemporal models is that they require dedicated software such as the WinBUGS package for Bayesian estimation and relatively long computational time and resources, especially for large datasets although this is becoming less of an issue with improving hardware and software implementations. The most common way to model spatial effects has been to use a Conditional Auto-Regressive (CAR) prior. However, similar to the case of KDE, the CAR prior imposes a uniform level of spatial smoothness (Lee, 2013) which may be too restrictive for crash data. It would be interesting to develop methods that allow the correlation structure to change resulting in different levels of smoothness from the CAR prior. The SMA could potentially provide insight on how to spatially model the correlation structure.

**Table 5 False Positive and MSPE**

| α | Reference Model | Evaluated Method | | |
|---|---|---|---|---|
| | | Count | EB | SMA |
| Top 1% | Count Model | **62.92%** | 64.17% | **62.92%** |
| | EB model | 61.52% | 60.42% | **59.58%** |
| | SMA Model | 60.83% | 62.08% | **58.33%** |
| Top 2.5% | Count Model | 56.43% | 53.92% | **49.75%** |
| | EB model | 54.92% | 50.42% | **47.75%** |
| | SMA Model | 54.76% | 52.09% | **46.41%** |
| Top 5% | Count Model | 48.83% | 46.61% | **43.24%** |
| | EB model | 48.50% | 41.99% | **40.73%** |
| | SMA Model | 46.66% | 42.99% | **36.81%** |
| Top 10% | Count Model | 44.84% | 39.87% | **36.16%** |
| | EB model | 42.76% | 33.32% | **31.57%** |
| | SMA Model | 40.58% | 32.61% | **25.72%** |
| MSPE | - | 1.8004 | 1.3781 | **1.2831** |

**Conclusions**

The SMA is a fast and accurate approach to estimate the expected number of crashes at road sections. It is similar in spirit to the KDE approach with the additional benefit of allowing different window sizes (bandwidth size) at different locations. One of the advantageous features of the SMA compared to the EB or full Bayes methods is that the SMA only uses crash counts to accurately estimate the expected number of crashes; there is no need to collect information about explanatory variables (not even traffic) as required by regression models (EB and full Bayes). However, sometimes some explanatory variables are readily available leading us to ask "Is it possible to include this additional information within the SMA approach or is it possible to include the SMA approach within the EB or full Bayes approach?"

In terms of including information from important variables, the authors are currently working on a method to incorporate information from crash modification factors into the SMA. Specifically, the authors are working on an approach to calculate Poisson's Unbiased Risk Estimate (PURE) for weighted crash counts (crash modification factors have a multiplicative effect on crash risk which suggests a weighted approach). Including the SMA into and EB or full Bayes approach seems more challenging although it would be an interesting topic for investigation.

In the SMA approach, a single threshold is optimized and used at each aggregation (decomposition) level. Because for Poisson data, the variance is equal to the mean, a potential improvement could be obtained if at each level, the threshold also depends on the variance (more specifically, the standard deviation). The true variance is not known however, an estimate of the variance from the data could be used to verify if this can lead in a better estimate of the expected number of crashes.

**Data and Software Availability**

In addition to the Matlab implementation given in the Appendix, an Excel implementation of the SMA with an example dataset can be obtained at the following link: https://github.com/johnsamer/Crash-Analysis-MHW-Sheet

**References**

1. Anderson, T.K. Kernel density estimation and K-means clustering to profile road accident hotspots. *Accidents Analysis and Prevention*, 2009. 41: 359–364.
2. Aguero-Valverde, J., & Jovanis, P. P. (2008). Analysis of road crash frequency with spatial models. *Transportation Research Record*, *2061*(1), 55-63.
3. Aguero-Valverde, J., & Jovanis, P. P. (2010). Spatial correlation in multilevel crash frequency models: Effects of different neighboring structures. *Transportation Research Record*, *2165*(1), 21-32.
4. Barua, S., El-Basyouny, K., & Islam, M. T. (2016). Multivariate random parameters collision count data models with spatial heterogeneity. *Analytic methods in accident research*, *9*, 1-15.
5. Bil, M., Andrasik, R., and Janoska Identification of hazardous road locations of traffic accidents by means of kernel density estimation and cluster significance evaluation. *Accident Analysis and Prevention*, 2013. 55: 265–273.
6. Cheng, W., and Washington, S.P. Experimental evaluation of hotspot identification methods. *Accident Analysis and Prevention*, 2005. 37: 870–881.
7. Cheng, W., & Washington, S. (2008). New criteria for evaluating methods of identifying hot spots. *Transportation Research Record: Journal of the Transportation Research Board*, (2083), 76-85.
8. Cheng, W., Gill, G. S., Dasu, R., Xie, M., Jia, X., & Zhou, J. (2017). Comparison of Multivariate Poisson lognormal spatial and temporal crash models to identify hot spots of intersections based on crash types. *Accident Analysis & Prevention*, *99*, 330-341.

9. Chung, K., Ragland, D.R., Madanat, S., and Oh, S. The continuous risk profile approach for the identification of high collision concentration locations on congested highways. *Transportation and Traffic Theory 2009: Golden Jubilee*, 2009. 463–480.
10. Dong, N., Huang, H., Lee, J., Gao, M., & Abdel-Aty, M. (2016). Macroscopic hotspots identification: a Bayesian spatio-temporal interaction approach. *Accident Analysis & Prevention*, *92*, 256-264.
11. El-Basyouny, K., & Sayed, T. (2009). Urban arterial accident prediction models with spatial effects. *Transportation Research Record: Journal of the Transportation Research Board*, (2102), 27-33.
12. Elvik, R. (2007). State-of-the-art approaches to road accident black spot management and safety analysis of road networks. Institute of Transport Economics, Oslo.
13. Elvik, R. A survey of operational definitions of hazardous road locations in some European countries. *Accident Analysis and Prevention*, 2008. 40: 1830–1835.
14. Elvik, R. The predictive validity of empirical Bayes estimates of road safety. *Accident Analysis and Prevention*, 2008. 40(6): 1964 – 1969.
15. Fawcett, L., Matthews, J., Thorpe, N., and Kremer, K. A full Bayes approach to road safety hotspot identification with prediction validation. *Presented at 97th the Annual Transportation Research Board Meeting*. 2018, Paper Number: 18-01020.
16. Flahaut, B., Mouchart, M., San Martin, E., and Thomas, I. The local spatial autocorrelation and the kernel method for identifying black zones: a comparative approach. *Accident Analysis and Prevention*, 2003. 35: 991–1004.
17. Freedman, D. From association to causation via regression. *Advances in Applied Mathematics*, 1997. 18(1): 59 – 110.
18. Gelman, A., and Hill, J. (2007). *Data analysis using regression and multilevel hierarchical models*. New York, NY, USA: Cambridge University Press.
19. Hauer, E. (1997). *Observational before/after studies in road safety. Estimating the effect of highway and traffic engineering measures on road safety*. Pargamon Press, Elsevier Science Ltd, Oxford, England.
20. Hauer, E., & Persaud, B. N. (1984). Problem of identifying hazardous locations using accident data. *Transportation Research Record*, *975*, 36-43.
21. Hauer, E. (2015). *The art of regression modeling in road safety*. New York: Springer.
22. Hauer, E., Harwood, D., Council, F., and Griffith, M. Estimating safety by the empirical Bayes method: a tutorial. *Transportation Research Record: Journal of the Transportation Research Board*, 2002. 1784: 126 – 131.
23. *Highway Safety Manual, Volume 2*. AASHTO, Washington, D.C., 2010.
24. Hirakawa, K., and Wolfe, P.J. Skellam shrinkage: Wavelet-based intensity estimation for inhomogeneous Poisson data. *IEEE Transactions on Information Theory*, 2012. 58(2): 1080 – 1093.
25. Huang, H., Chin, H., & Haque, M. (2009). Empirical evaluation of alternative approaches in identifying crash hot spots: naive ranking, empirical Bayes, and full Bayes methods. *Transportation Research Record: Journal of the Transportation Research Board*, (2103), 32-41.
26. Huang, H., Song, B., Xu, P., Zeng, Q., Lee, J., & Abdel-Aty, M. (2016). Macro and micro models for zonal crash prediction with application in hot zones identification. *Journal of Transport Geography*, *54*, 248-256.
27. Jia, R., Khadka, A., & Kim, I. (2018). Traffic crash analysis with point-of-interest spatial clustering. *Accident Analysis & Prevention*, *121*, 223-230.

28. Katicha, S.W., and Flintsch, G.W. Multiscale Vehicular Expected Crashes Estimation with the Unnormalized Haar Wavelet Transform and Poisson's Unbiased Risk Estimate. *Journal of Transportation Engineering, Part A: Systems*, 2018. 144(8): 04018037.
29. Kweon, Y., and Lim, I. (2014). Development of Safety Performance Functions for Multilane Highway and Freeway Segments Maintained by the Virginia Department of Transportation. Final Report VCTIR 14-R14, Charlottesville, Virginia.
30. Lee, D. (2013). CARBayes: an R package for Bayesian spatial modeling with conditional autoregressive priors. *Journal of Statistical Software*, 55(13), 1-24.
31. Lord, D., & Park, P. Y. J. (2008). Investigating the effects of the fixed and varying dispersion parameters of Poisson-gamma models on empirical Bayes estimates. *Accident Analysis & Prevention*, 40(4), 1441-1457.
32. Luisier, F., Vonesch, C., Blu, T., and Unser, M. Fast interscale denoising of Poisson-corrupted images. *Signal Processing*, 2010. 90(2): 415 – 427.
33. Loo, B. P., Yao, S., & Wu, J. (2011, June). Spatial point analysis of road crashes in Shanghai: A GIS-based network kernel density method. In *2011 19th international conference on geoinformatics* (pp. 1-6). IEEE.
34. Mannering, F.L., and Bhat, C.R. Analytic methods in accident research: Methodological frontiers and future directions. *Analytic methods in accident research*, 2014. 1: 1 – 22.
35. Miaou, S. P., & Song, J. J. (2005). Bayesian ranking of sites for engineering safety improvements: decision parameter, treatability concept, statistical criterion, and spatial dependence. *Accident Analysis & Prevention*, 37(4), 699-720.
36. Montella, A. A comparative analysis of hotspots identification methods. *Accident Analysis and Prevention*, 2010. 42: 571–581.
37. Okabe, A., Satoh, T., and Sugihara, K. A kernel density estimation method for networks, its computational method and GIS-based tool. *International Journal of Geographical Information Science*, 2009. 23:1, 7-32.
38. O'Sullivan, D., & Wong, D. W. A surface-based approach to measuring spatial segregation. *Geographical Analysis*, 2007, 39(2), 147-168.
39. Park, B., Lord, D., and Lee, C. Finite mixture modeling for vehicle crash data with application to hotspot identification. *Accident Analysis and Prevention*, 2014. 71: 319–326.
40. Persaud, B., Lan, B., Lyon, C., and Bhim, R. Comparison of empirical Bayes and full Bayes approaches for before-after road safety evaluations. *Accident Analysis and Prevention*, 2010. 42: 38 – 43.
41. Qu, X., and Meng, Q. A note on hotspot identification for urban expressways. *Safety Science*, 2014. 66: 87–91.
42. Jia, R., Khadka, A., & Kim, I. (2018). Traffic crash analysis with point-of-interest spatial clustering. *Accident Analysis & Prevention*, 121, 223-230.
43. Silverman, B.W. (2018), *Density Estimation for Statistics and Data Analysis*. Routledge.
44. Thomas, I. (1996). Spatial data aggregation: exploratory analysis of road accidents. *Accident Analysis & Prevention*, 28(2), 251-264.
45. Woodward, J. (2003). *Making things happen: a theory of causal explanation*. Oxford university press.
46. Xie, Z., and Yan, J. Kernel density estimation of traffic accidents in network space. *Computers, Environment and Urban Systems*, 2008. 32: 396–406.
47. Yu, H., Liu, P., Chen, J., and Wang, H. Comparative analysis of the spatial analysis methods for hotspot identification. *Accident Analysis and Prevention*, 2014. 66: 80–88.

## Appendix

The following Matlab function implements the SMA approach:

```matlab
function Crash_Risk = SMA(Crash_Counts)
NumLevels = floor(log2(length(Crash_Counts)));
[SUMS, DIFFERENCES] = Calculate_SUMS_DIFFERENCES(Crash_Counts,NumLevels);
T_DIFFERENCES = Threshold_DIFFERENCES(SUMS,DIFFERENCES);
Crash_Risk = Estimate_Crash_Risk(SUMS,T_DIFFERENCES);
%%%%%%%%%%%%%%%%%%%%%%%%%%%%%%%%%%%%%%%%%%%%%%%%%%%%%%%%%%%%%%%%%%%%%%%%%%%
%%%%%%%%%%%%%%%%%%%%%%%%%
%% Main Functions
%%% Calculate_SUMS_DIFFERENCES
function [SUMS,DIFFERENCES] = Calculate_SUMS_DIFFERENCES(Crash_Counts,Levels)
Num_Of_Sections = length(Crash_Counts);
SUMS = zeros(Num_Of_Sections,Levels);
DIFFERENCES = zeros(Num_Of_Sections,Levels);
SUMS(:,1) = (Crash_Counts+shift(Crash_Counts,-1));
DIFFERENCES(:,1) = (Crash_Counts-shift(Crash_Counts,-1));
for i=2:Levels
    SUMS(:,i) = (SUMS(:,i-1)+shift(SUMS(:,i-1),-2^(i-1)));
    DIFFERENCES(:,i) = (SUMS(:,i-1)-shift(SUMS(:,i-1),-2^(i-1)));
end
%%% Threshold_DIFFERENCES
function T_DIFFERENCES = Threshold_DIFFERENCES(SUMS,DIFFERENCES)
Levels = size(DIFFERENCES,2);
T_DIFFERENCES = zeros(size(DIFFERENCES));
for i=1:Levels
    TH = Determine_Threshold(SUMS(:,i),DIFFERENCES(:,i));
    T_DIFFERENCES(:,i) = Threshold(DIFFERENCES(:,i),TH);
end
%%% Estimate_Crash_Risk
function Crash_Risk = Estimate_Crash_Risk(SUMS,T_DIFFERENCES)
Levels = size(T_DIFFERENCES,2);
Crash_Risk = (SUMS(:,end)+T_DIFFERENCES(:,end)+shift((SUMS(:,end)-
T_DIFFERENCES(:,end)),2^(Levels-1)))/2/2;
for i=Levels-1:-1:1
    Crash_Risk = max((Crash_Risk+T_DIFFERENCES(:,i)+shift((Crash_Risk-
T_DIFFERENCES(:,i)),2^(i-1)))/2/2,0);
end
%%%%%%%%%%%%%%%%%%%%%%%%%%%%%%%%%%%%%%%%%%%%%%%%%%%%%%%%%%%%%%%%%%%%%%%%%%%
%%%%%%%%%%%%%%%%%%%%%%%%%
%% Secondary Functions
%%% Determine_Threshold
function TH = Determine_Threshold(S,D)
n = length(S);
NumOfTestThreshold = 40;
TH_test = linspace(0,max(sqrt(S))*sqrt(8*log(n)),NumOfTestThreshold);
PURE_Profile = zeros(NumOfTestThreshold,1);
for i=1:NumOfTestThreshold
    PURE_Profile(i) = PURE(S,D,TH_test(i));
end
[~,id] = min(PURE_Profile);
TH = TH_test(id(1));
%%% Threshold
function T_D = Threshold(D,th)
s = sign(D);
```

```matlab
T_D = s.*max(abs(D).*(1-(th./abs(D)).^2),0);
%%% PURE
function P = PURE(S,D,TH_test)
F1 = Threshold(D,TH_test)-D;
F2 = Threshold(D-1,TH_test)-(D-1);
F3 = Threshold(D+1,TH_test)-(D+1);
P = sum(S+F1.^2+2*D.*F1-(S+D).*F2+(S-D).*F3);
%%% shift
function y = shift(x,shift_size)
id = 1:length(x);
if shift_size>0
    id = [id(end-shift_size+1:end) id(1:end-shift_size)];
elseif shift_size<=0
    id = [id(1-shift_size:end) id(1:-shift_size)];
end
y = x(id);
```